\documentclass[prl,twocolumn,showpacs,superscriptaddress,preprintnumbers,amsmath,amssymb]{revtex4-2}
\usepackage{hyperref}
\usepackage{amsmath,amssymb,bm}
\usepackage[dvips]{graphicx}
\usepackage{yfonts}
\usepackage{enumerate}
\usepackage{color}
\usepackage{url}

\newcommand{\rank}{\mathop{\mathrm{rank}}}
\newcommand{\ri}{\mathrm{i}}

\begin{document}
\preprint{KEK-TH-2256, J-PARC-TH-0226}
\title{Rainbow Nambu-Goldstone modes  under a shear flow}

\author{Yuki Minami}
\affiliation{Department of Physics, Zhejiang University, Hangzhou 310027, China}
\email{yminami@zju.edu.cn}
\author{Hiroyoshi Nakano}
\affiliation{Department of Physics, Kyoto University, Kyoto 606-8502, Japan}
\email{h.takagi@scphys.kyoto-u.ac.jp}
\author{Yoshimasa Hidaka}
\affiliation{KEK Theory Center, Tsukuba 305-0801, Japan}
\affiliation{Graduate University for Advanced Studies (Sokendai), Tsukuba 305-0801, Japan}
\affiliation{RIKEN iTHEMS, RIKEN, Wako 351-0198, Japan}
\email{hidaka@post.kek.jp}

\begin{abstract}
We study an $O(N)$ scalar model under shear flow and
its Nambu-Goldstone modes associated with spontaneous symmetry breaking $O(N) \to O(N-1)$.
We find that the Nambu-Goldstone mode splits into an infinite number of gapless modes, which we call the rainbow Nambu-Goldstone modes. They have different group velocities and the fractional dispersion relation $\omega \sim k_1^{2/3}$, where $k_1$ is the wavenumber along the flow.
Such behaviors do not have counterparts in an equilibrium state. 
\end{abstract}

\maketitle

\emph{Introduction.}---
Gapless modes appear universally in various systems, and they govern macroscopic behaviors~\cite{chaikin1995principles, kittel1996introduction}.
For example, phonons, which are the gapless modes in a solid crystal, have the liner dispersion $\omega \sim k$, where $\omega$ is the frequency, and $k$ is the wavenumber. It leads to the Debye $T^3$ law of the specific heat at low temperature~\cite{kittel1996introduction}. 
Similarly, magnons in a ferromagnet, which have $\omega \sim k^2$, explain the temperature dependence of magnetization known as the Bloch $T^{3/2}$ law~\cite{PhysRev.132.2051}.
The role of gapless modes becomes more significant in lower dimensions; they change the infrared behaviors: The magnons cause the infrared divergence of spin fluctuations in two dimensions at finite temperature, and it breaks the magnetization~\cite{chaikin1995principles}.
Shear and heat diffusion modes in a fluid, $\omega \sim -\ri k^2$, lead to a long-time tail of a correlation function and divergences of the thermal conductivity and the shear viscosity in a two dimensional system~\cite{PhysRevA.16.732, Kawasaki}.
Therefore, it is important to study  behaviors of the gapless modes.

An important class of gapless modes is due to spontaneous symmetry breaking.
The Nambu-Goldstone (NG) theorem~\cite{nambu1961dynamical, goldstone1961field, goldstone1962broken} shows the existence of gapless modes when a continuous symmetry is spontaneously broken.
Symmetry also restricts the number of NG modes and their dispersions as well as the interactions~\cite{nambu1961dynamical, goldstone1961field, goldstone1962broken, Watanabe:2012hr, Hidaka:2012ym, Watanabe:2014fva, Hayata:2014yga, Takahashi:2014vua, Minami:2018oxl, hidaka2020spontaneous}.
For an isolated system without the Lorentz symmetry, when a continuous internal symmetry is spontaneously broken, the number of NG modes is generally expressed as~\cite{Watanabe:2012hr, Hidaka:2012ym, Watanabe:2014fva,Hayata:2014yga,Takahashi:2014vua}
\begin{align}
N_\text{NG}=N_\text{BS} - \frac{1}{2}\rank\rho,
\label{eq:NGRelation}
\end{align}
where $N_\text{BS}$ is the number of broken symmetries or generators. $\rho^{\beta\alpha}:=-\langle[\ri {Q}^\beta,{j}^{\alpha 0}(x)]\rangle$ is a $N_\text{BS}\times N_\text{BS}$ matrix, where $Q^{\beta}$ and $j^{\alpha0}(x)$ are the Noether charge and its charge density, respectively~\cite{Watanabe:2011ec}. The indices $\alpha$ and $\beta$ run over the index of broken generators. The charges belonging to the kernel of $\rho$ are called type-A, whose number is equal to $N_\text{BS}-\rank \rho=:N_\text{A}$. 
The others are called type-B.
The type-B charges form $(\rank\rho)/2=:N_\text{B}$ canonical pairs, and each pair corresponds to each type-B NG mode.
The sum of $N_\text{A}$ and $N_\text{B}$ is equal to the number of NG modes, which satisfies Eq.~\eqref{eq:NGRelation}~\cite{ Watanabe:2012hr, Hidaka:2012ym}. The dispersion relations of type-A and type-B NG modes are typically $\omega\sim k$ and $\omega \sim k^2$, respectively. 
This classification can be generalized for symmetries of extended objects called higher-form symmetries~\cite{gaiotto2015generalized,Lake:2018dqm,Hidaka:2020ucc}.

Furthermore, the NG modes in open systems, where conservation laws are violated by interactions with an environment, are recently discussed in terms of spontaneous symmetry breaking~\cite{Minami:2018oxl, hidaka2020spontaneous,Hayata:2018qgt,Hongo:2019qhi, baggioli2019zoology, baggioli2020homogeneous, baggioli2020effective}. 
An example of NG modes in an open system is an exciton-polariton in a microcavity.
The exciton of a semiconductor strongly couples to cavity photons and it causes a new bosonic mode considered as a NG mode~\cite{carusotto2013quantum, 0034-4885-79-9-096001}.  
The NG mode associated with $U(1)$ breaking is 
diffusive~\cite{szymanska2006nonequilibrium, wouters2007excitations}, which is not propagating and not classified into the types of NG modes in isolated systems.

The Nambu-Goldstone theorem has been extended to open systems in Ref.~\cite{Minami:2018oxl,hidaka2020spontaneous}.
The Noether charges double in the path-integral formalism for open systems~~\cite{Minami:2018oxl}.
Using the doubled charges, the NG modes are classified into four types: type-A propagation, type-A diffusion, type-B propagation, and type-B diffusion modes, and  the relation between these NG modes and broken symmetries is shown~\cite{hidaka2020spontaneous}. Although the obtained results are general and widely applicable, it is limited in homogenous systems where translational symmetry is not broken.

One of the interesting generalizations is to consider NG modes in a system under a nonequilibrium steady flow, which explicitly breaks the translational symmetry.
In such a system, we expect that a new type of NG mode appears.
A known interesting phenomenon specific to nonequilibrium steady states
is long-range correlations, which are absent in equilibrium states~\cite{dorfman1994generic, OnukiKawasaki}.
In a three dimensional fluid with a constant temperature gradient $\nabla T$, the spatial correlation of the density fluctuation behaves as~\cite{PhysRevA.26.1812, dorfman1994generic}
\begin{align}
S_{nn}(\bm{k})&=\int d\bm{x} e^{-\ri \bm{k}\cdot \bm{x}}
\langle \delta n(\bm{x}) \delta n(\bm{0})
\rangle_{\rm NESS}^T, \notag \\
&\sim S_{\rm eq}(\bm{k})\Bigl(1+\frac{(\hat{\bm{k}}_\perp \cdot \nabla T)^2}{k^4}\Bigr),
\label{heat flow}
\end{align}
where $\langle ....
\rangle_{\rm NESS}^T$ is the average in the steady state under the temperature gradient, and $\hat{\bm{k}}_\perp$ is the unit vector perpendicular to $\bm{k}$. $ S_{\rm eq}(\bm{k})$ is the correlation in the equilibrium state without the temperature gradient.
The temperature gradient enhances the long-range correlation at $k \to 0$,
and it is singular.
Similarly, in a three dimensional critical fluid under shear flow,
the spatial correlation of the entropy fluctuation $\delta s$, which is the order parameter of the liquid-gas transition, behaves as~\cite{OnukiKawasaki}
\begin{align}
\int d\bm{x} e^{-\ri \bm{k}\cdot \bm{x}}
\langle \delta s(\bm{x}) \delta s(\bm{0})
\rangle_{\rm NESS}^{\rm s} 
\sim |k_1|^{-2/5},
\end{align}
where $\langle ...\rangle_{\rm NESS}^{\rm s} $ is the average in the steady state under the shear flow and $k_1$ is the wavenumber along the flow.
The spatial correlation has the fractional behavior, and is singular at $k_1 \to 0$.

These results indicate that nonequilibrium steady flows drastically change the behavior of the gapless modes.
In this letter, we study behaviors of NG modes under a nonequilibrium steady flow by using a toy model.
We shall find that the NG mode splits into an infinite number of gapless modes, which we call the rainbow NG modes; they have the fractional dispersion relation $\omega \sim k_1^{2/3}$ and different group velocities. To the best of our best knowledge, such behaviors do not have counterparts in equilibrium states.

\begin{figure}[htbp]
  \begin{center}
    \begin{tabular}{c}
      \begin{minipage}{0.4\hsize}
        \begin{center}
          \includegraphics[width=\hsize]{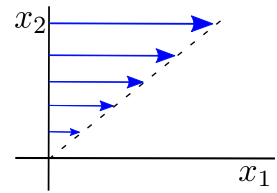}
        \end{center}
      \end{minipage} 
    \end{tabular}
    \caption{The blue arrow shows the velocity of shear flow.
    The flow is along the $x_1$ axis, and the velocity has the constant gradient along the $x_2$ axis. 
    }
    \label{fig:shear}
  \end{center}
\end{figure}

\emph{Model.}---
We consider a Langevin equation of an $O(N)$ scalar model under a flow in $(3+1)$ spacetime dimensions~\cite{de1976effect, onuki1980critical, haga2015nonequilibrium},
\begin{align}
\partial_t \phi^a+{(\bm{v}\cdot\nabla)} \phi^a&=-\Gamma \frac{\delta F}{\delta \phi^a}+\eta^a, 
\label{eq:modelA}
\end{align}
where $\phi^a(t, \bm{x})$ is the $N$-component real scalar fields, $\bm{v}(\bm{x})$  the steady flow velocity, $\Gamma$ the diffusion constant, and $F$ the free energy. $\eta(t,\bm{x})$ represents the noise satisfying the fluctuation dissipation relation
\begin{align}
  \langle \eta^a(t,\bm{x})\eta^b(t',\bm{y})\rangle&=2T\Gamma\delta^{ab}\delta(t-t')\delta(\bm{x}-\bm{y}),
  \label{eq:noise}
\end{align}
where $T$ is the temperature. 
As the flow $\bm{v}(\bm{x})$, we consider the shear flow shown in Fig.~\ref{fig:shear}, 
\begin{align}
\bm{v}(\bm{x}) = (\gamma x_2, 0, 0),
\end{align}
where $\gamma$ represents the magnitude of the velocity gradient, and we assume $\gamma >0$.
Here, we suppose that $x_2$ direction has the boundary at $x_2=0$, i.e., the fields $\phi^a$ are located on $0 \leq x_2 < \infty$, while the other directions have no boundary, $-\infty < x_1, x_3 < \infty$. 
We assume that the free energy $F$ has the form
\begin{align}
F=\int d\bm{x}\Bigl[ \frac{1}{2}(\nabla \phi^a)^2-\frac{\mu^{2}}{2}(\phi^a)^2+\frac{u^2}{4}((\phi^a)^2)^2\Bigr],\label{eq:free energy}
\end{align}
where $\mu$ and $u$ are parameters, and the Einstein summation convention is adopted. Without loss of generality, we take that $\mu$ and $u$ are positive valued. The equation of motion under the shear flow is written as
\begin{align}
  \!\!\!\!\Bigl(\partial_t+\gamma x_2 \frac{\partial}{\partial x_1}\Bigr)\phi^a =\Gamma\bigl((\nabla^2+\mu^2)
  \!-\!u^2(\phi^b)^2\bigr)\phi^a +\eta^a.
\label{eq:modelA}
\end{align}
%The time-reversibility of $\gamma x_2 \partial_{x_1} \phi^a$ and $\bigl((\nabla^2+\mu^2)  \!-\!u^2(\phi^b)^2\bigr)\phi^a $ are even and those are irreversible processes.
This equation and the fluctuation dissipation relation~\eqref{eq:noise} are covariant under $\phi^a\to R^a_{~b}\phi^b$ and $\eta^a\to R^{a}_{~b}\eta^{b}$, and thus, the equation has $O(N)$ symmetry. Here, $R^a_{~b}$ is a rotational matrix.

Let us find the steady state solutions within the mean field approximation. They are obtained as 
the homogeneous saddle points of the free energy
\begin{align}
\left.\frac{\delta F}{\delta \phi^a}\right|_{\phi^a=\bar{\phi}^a} = (-\mu^2+u^2(\bar{\phi}^b)^2)\bar{\phi}^a=0.
\end{align}
There are two types of solutions: One is the trivial solution, $\bar{\phi}^a=0$. 
However, the trivial solution is not favored because it is unstable against a small fluctuation,
$\delta\phi^a\sim e^{\Gamma \mu^2 t}$. The other satisfies $(\bar{\phi}^a)^2=\mu^2 / u^2$. 
This equation determines only the radius of $\phi^a$, so that the solutions are degenerate on $(N-1)$-dimensional surface. Since these solutions are equivalent, we may choose $\bar{\phi}^a=(\mu/u,\bm{0})$ as the solution. 
Obviously, the solution is not invariant under a general $O(N)$ rotation, but still invariant under $O(N-1)$ transformation that does not rotate the first component. 
Therefore, $O(N)$ is spontaneously broken into $O(N-1)$.

The Nambu-Goldstone theorem implies that there are gapless modes as fluctuations around the steady state~\cite{nambu1961dynamical, goldstone1961field, goldstone1962broken, Minami:2018oxl, hidaka2020spontaneous}. To see this, we expand the field variables as $\phi^a(t,\bm{x})=\bar{\phi}^a+(\sigma(t,\bm{x}),\pi^b(t,\bm{x}))$,
where the suffix runs $b=2,3,..,N$.
In this parametrization, the Langevin equations for $\sigma$ and $\pi^b$ are expressed as 
\begin{align}
\bigl(\partial_t+\gamma x_2 \partial_{x_1}-\Gamma  \nabla^2+2\Gamma\mu^2\bigr)\sigma=&\eta^1+\cdots,  \label{sigma} \\
\bigl(\partial_t+\gamma x_2 \partial_{x_1}-\Gamma \nabla^2\bigr)\pi^b=&\eta^b+\cdots, \label{NG}
\end{align}
where the last terms ``$\cdots$'' represent the nonlinear terms. 
Because there is the nonvanishing term $2\Gamma \mu^2 \sigma$ on the left-hand side of Eq.~(\ref{sigma}) for a constant $\sigma$, it is gapped. 
In contrast, the left-hand side of Eq.~(\ref{NG}) vanishes for a constant $\pi^b$; i.e., $\pi^b$ are gapless.

\emph{Linearized Langevin equation.}---
\label{sec:Airy}
To derive the dispersion relation of $\pi^b$, we need to solve the linearized equation for fluctuations, which we obtain by dropping the nonlinear term in Eq.~(\ref{NG}).
Performing the Fourier transform for $t$, $x_1$ and $x_3$, 
we obtain the following linearized Langevin equation;
\begin{align}
\Bigl(-\ri\omega&+\ri \gamma x_2 k_1+\Gamma {\bm{k}_\perp^2}-\Gamma\frac{\partial^2}{\partial x_2^2}\Bigr)\pi^b(x_2,{\bm{k}_\perp},\omega)
\nonumber \\
&=\eta^b(x_2,{\bm{k}_\perp},\omega). \label{eigeneq}
\end{align}
Here, $\omega$ is the frequency and we define $\bm{k}_\perp:=(k_1,k_3)$.
It is useful to introduce dimensionless parameters as
$\bm{X}=\sqrt{\gamma/\Gamma}\bm{x}$,  
$\bm{K}=\sqrt{\Gamma /\gamma}\bm{k}$, and   
$\Omega=\omega/\gamma$.
Then, the equation for $\pi^b$ becomes
\begin{align}
\Bigl(-\frac{\partial^2}{\partial X_2^2}&+\ri K_1X_2-\ri\Omega+{\bm{K}_\perp}^2\Bigr)\pi^b(X_2,{\bm{K}_\perp},\Omega) \nonumber \\
&=\zeta^b(X_2,{\bm{K}_\perp},\Omega). \label{eq:NGEoM}
\end{align}
Here, we define $\pi^b(X_2,{\bm{K}_\perp},\Omega):=\pi^b(x_2,{\bm{k}_\perp}\omega)$, and the noise $\zeta^b(X_2,{\bm{K}_\perp},\Omega) := \eta^b(x_2,{\bm{k}_\perp},\omega)/\gamma$.
To solve Eq.~(\ref{eq:NGEoM}), we consider the following operator and its eigenvalue equation;
\begin{align}
&\hat{\mathcal{L}}(K_1)=-\frac{\partial^2}{\partial X_2^2}+\ri K_1X_2, \label{eq:opL} \\
&\hat{\mathcal{L}}(K_1) \psi_n(X_2,K_1)=\lambda_n(K_1)\psi_n(X_2,K_1), \label{eq:eigen eq} 
\end{align}
where $\psi_n$ is the eigenfunction, and  $\lambda_n$ is the eigenvalue.
We impose the boundary condition such that the fluctuation vanishes at the boundary and the infinity, 
\begin{align}
  \psi_n(0,K_1)=0 \quad\text{and} \lim_{X'_2\to\infty}\psi_n(X'_2,K_1)=0. \label{eq:boundaryCondition}
\end{align}
The operator (\ref{eq:opL}) is called the complex Airy operator~\cite{Savchuk2017}, which is non-Hermitian, but complex symmetric.
By writing the eigenvalue equation as
\begin{align}
\Bigl[\frac{\partial^2}{\partial X_2^2}-\ri K_1X_2+\lambda_n(K_1)\Bigr]\psi_n(X_2,K_1)=0,
\end{align}
and changing the variable as $Y=(\ri K_1)^{1/3}X_2-(\ri K_1)^{-2/3}\lambda_n(K_1)$,
we obtain $(\partial^2/\partial Y^2- Y)\psi_n(Y)=0$.
That is, the eigenvalue equation turns to  the Airy equation~\cite{Airy}.
The general solution is written by the Airy functions Ai$(Y)$ and Bi$(Y)$ as
$\psi_n(Y)=a_n\, {\rm Ai}(Y)+b_n\, {\rm Bi}(Y)$,
where $a_n$ and $b_n$ are superposition coefficients.
 
Now, we determine the eigenvalues and the eigenfunctions by imposing the boundary condition~\eqref{eq:boundaryCondition}. 
First, we consider the condition at infinity.
An asymptotic behavior of the Airy function depends on the argument of $Y$ at $|Y|\to\infty$. In our model, it is calculated as
$|\arg(Y)|  \sim |\arg ((\ri K_1)^{1/3}X_2)|
=  \pi/6$.
From the asymptotic formulas for $|\arg(Y)| < \pi/3$,  ${\rm Ai}(Y) \to 0$, ${\rm Bi}(Y) \to \infty$ at $|Y| \to \infty$~\cite{Airy}, and thus, the coefficient $b$ must vanish.
Next, we consider the condition at $X_2=0$. It leads to
${\rm Ai}(Y_0) =0$ with $Y_0  := -(\ri K_1)^{-2/3}\lambda_n(K_1)$,
so that $Y_0$ have to be equal to zeros of Ai$(x)$. The zeros are simple and only located on the negative real axis.
Denoting $n$-th zero as $-t_n$, we find the eigenvalues and eigenfunctions as
\begin{align}
\lambda_n(K_1)&=(\ri K_1)^{2/3}t_n, \label{eigen value}\\
\psi_n(X_2,K_1)&= \mathcal{N}_n(K_1){\rm Ai}((\ri K_1)^{1/3} X_2-t_n), \label{eigen func}
\end{align}
where $\mathcal{N}_n(K_1)$ is the normalization constant and defined 
such that $ \int dX_2 |\psi_n|^2=1$.

\begin{figure}[htbp]
  \begin{center}
    \begin{tabular}{c}
      \begin{minipage}{0.5\hsize}
        \begin{center}
          \includegraphics[width=\hsize]{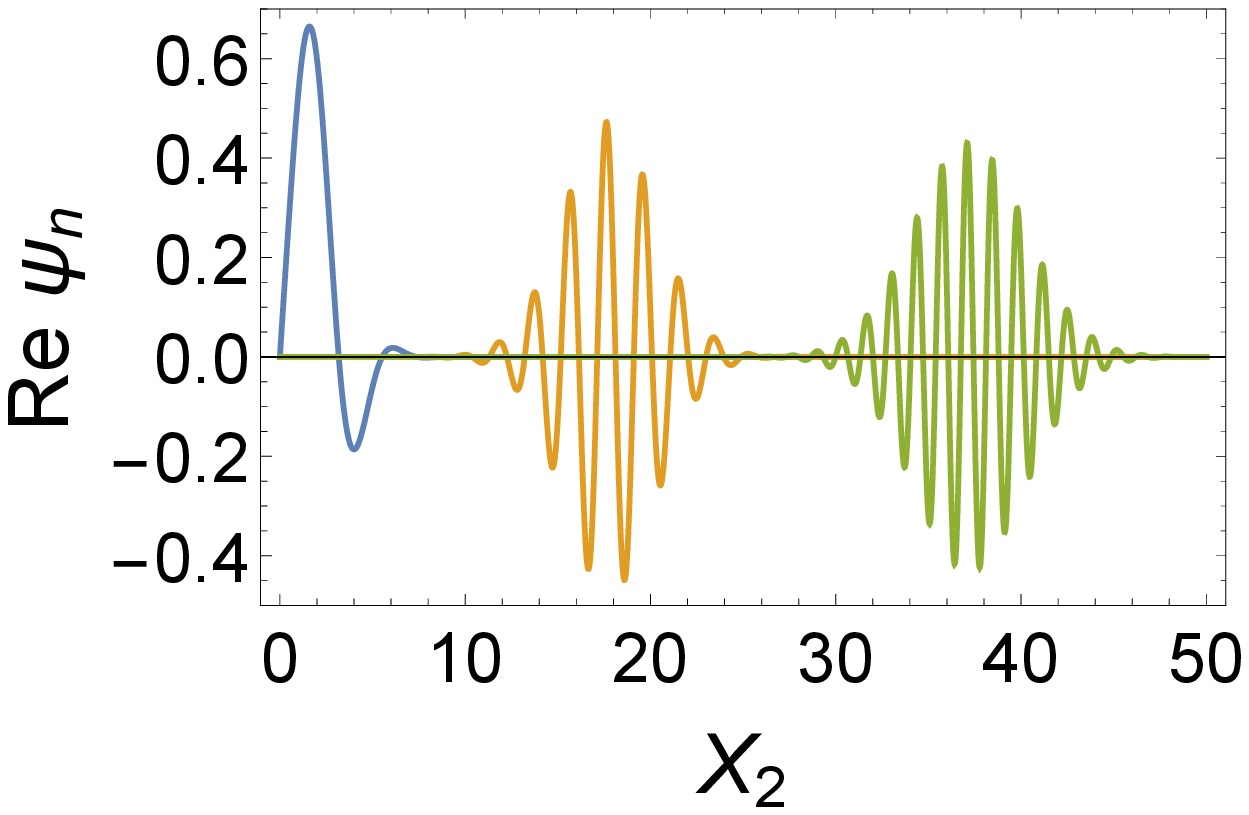}
        \end{center}
      \end{minipage} 
            \begin{minipage}{0.5\hsize}
        \begin{center}
          \includegraphics[width=\hsize]{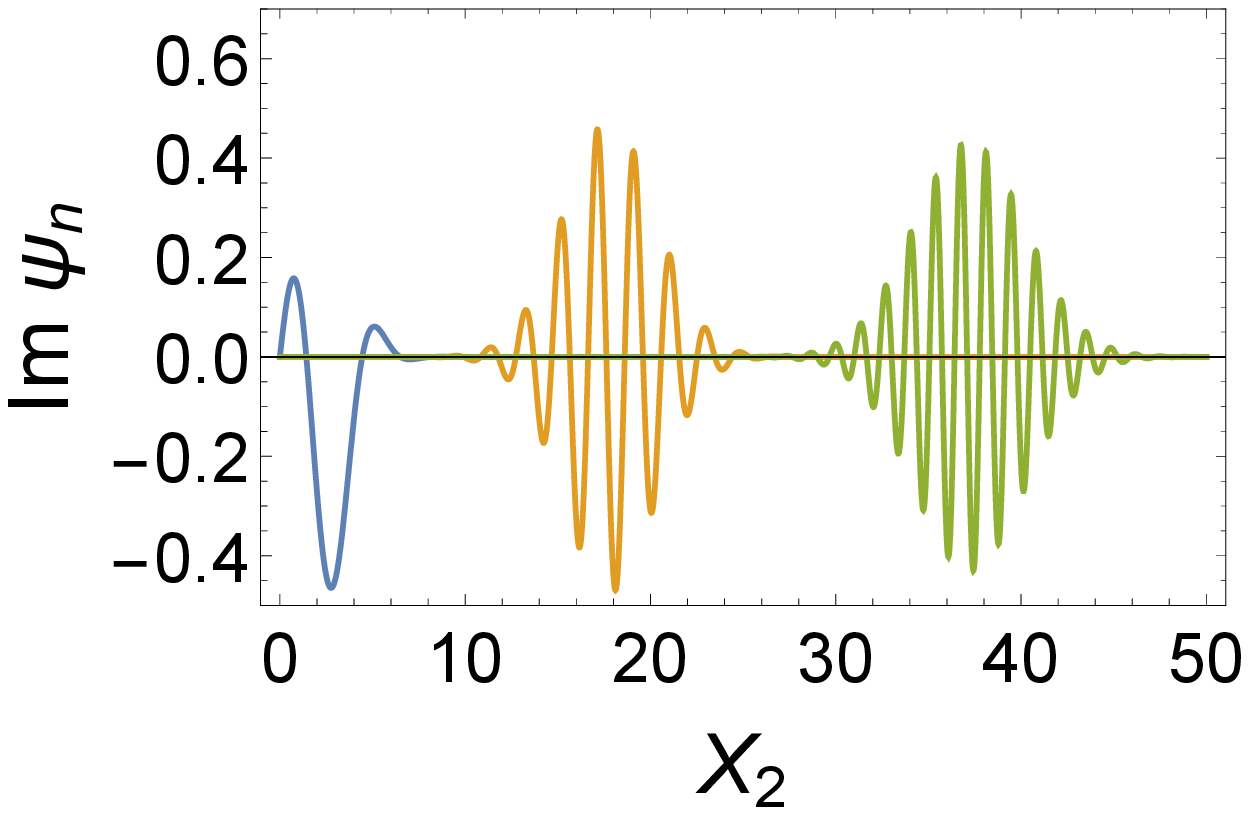}
        \end{center}
      \end{minipage} 
    \end{tabular}
    \caption{The real and imaginary parts of the eigenfunctions. The blue, orange,  and green  lines show the eigenfunctions at $n=1$, $20$, $60$, respectively. We have set $K_1=1$.}
    \label{fig:eigenfunction}
  \end{center}
\end{figure}
Figure~\ref{fig:eigenfunction} shows the real and the imaginary parts of the eigenfunctions for several $n$. 
The blue, orange, and green  lines show the eigenfunctions at $n=1$, $20$, $60$, respectively.
The eigenfunctions are localized oscillating functions.

It is shown that the eigenfunction system forms a complete system and  spans an orthogonal basis~\cite{Savchuk2017}.
We can expand an arbitrary square-integrable function as
\begin{align}
f(X_2,K_1) &=\sum_{n=1}^\infty f_n(K_1) \psi_n(X_2,K_1), \\
f_n(K_1) &=\int^\infty_0 \frac{dX_2}{\mathcal{C}_n(K_1)} \psi_n(X_2, K_1)f(X_2, K_1),
\end{align}
where $\mathcal{C}_n(K_1) = \int_0^\infty dX_2 \psi_n^2(X_2, K_1)$.
Expanding Eq.~(\ref{eq:NGEoM}) by Eq.~(\ref{eigen func}), we obtain the diagonal representation of the linearized Langevin equation,
\begin{align}
G^{-1}_n(\Omega, K_1, K_3)&\pi_n^b(K_1, \Omega)=\zeta^{b}_n(K_1,\Omega), \label{diagonal EOM} \\
G_n(\Omega, K_1, K_3)&=1/(\lambda_n(K_1)-i\Omega+K_1^2+K_3^2). \label{response function}
\end{align}
Here, we introduced the response function $G_n(\Omega, K_1, K_3)$ which is an observable in experiments.

\emph{Dispersion relation of NG modes.}---
The dispersion of NG modes can be obtained as the pole of the response function (\ref{response function}), which is
\begin{align}
\Omega&=-\ri(\ri K_1)^{2/3}t_n-\ri(K_1^2+K_3^2), \label{dispersion}\\
&=  \frac{\sqrt{3}}{2}{\rm sgn}(K_1)t_n|K_1|^{2/3}-\frac{\ri}{2}|K_1|^{2/3}t_n-\ri(K_1^2+K_3^2).\notag
\end{align}
Here, sgn$(x)$ is the sign function, and we decomposed it into the real and imaginary parts by using $(\pm \ri)^{2/3}=1/2\pm \ri\sqrt{3}/2$.
In the original variables, the dispersion relation becomes
\begin{align}
\omega=&  \frac{\sqrt{3}}{2}{\rm sgn}(\gamma k_1)\Gamma^{1/3}t_n|\gamma k_1|^{2/3}-\frac{\ri}{2}\Gamma^{1/3}t_n|\gamma k_1|^{2/3} \nonumber \\
&-\ri\,\Gamma(k_1^2+k_3^2).
\end{align}
This is the main result of this letter. There are several remarkable features:
First, the dispersion has the novel fractional exponent $2/3$.
Gapless modes with fractional dispersion themselves are not surprising. 
Such modes often appear if a spatial symmetry is broken, e.g., on the interface or domain-wall of a matter~\cite{Watanabe:2014zza}. A typical example is the gravity wave that is a surface wave on a fluid under gravity~\footnote{The readers should not confuse the gravity waves with gravitational waves appeared in general relativity.}. It has a fractional dispersion, $\omega \sim k^{1/2}$~\cite{landau1959lifshitz}.
Another example is the wave on an interface between two phase-separated Bose-Einstein condensates, which shows $\omega \sim k^{3/2}$~\cite{PhysRevA.65.033618}.
These power behavior can be understood as follows:
The equation of motion for a small fluctuation form a second order differential equation, and then the dispersion relation is obtained by solving the quadratic equation with respect to $\omega$. If the coefficients of the quadratic equation are polynomials of $k$, the dispersion relation behaves like $\omega \sim k^{n/2}$, with some integer $n$.

In this sense, the power of $2/3$ is somewhat unusual. Nevertheless, we can understand the power from the scale analysis:
Since $\omega \sim \lambda_n$ at small $k_1$ and $k_3$, the behavior is determined by the complex Airy operator $\hat{\mathcal{L}}$.
Let $\Delta X_2$ be the typical length scale of eigenfunctions.
Then, $X_2$ and $\partial/\partial X_2$ in $\hat{\mathcal{L}}$ behave like $X_2\sim \Delta X_2$ and $\partial/\partial X_2\sim 1/\Delta X_2$, respectively.
Therefore, the complex Airy operator is $\hat{\mathcal{L}}\sim (\Delta X_2)^{-2} -\ri K_{1}\Delta X_2$.
The first and second term must be balanced for the eigenstate, so that we obtain the typical scale $\Delta X_2\sim K_{1}^{-1/3}$ for given $K_{1}$. This leads to $\omega \sim \hat{\mathcal{L}}\sim K_{1}^{2/3}$ as we expected.

Second is the existence of a real part in Eq.~\eqref{dispersion}, which means that the NG modes are propagation modes.
 The system that we have considered is an open system in the sense that the system described by $\phi^a$ couples with the thermal bath. The NG modes in such a system are typically diffusive~\cite{Minami:2018oxl, hidaka2020spontaneous}. In fact, if we take $\gamma\to 0$, the dispersion of NG modes is $\omega\sim -\ri k^2$.
The shear flow gives a new propagation mechanism of NG modes.

Third is the group velocity of NG modes. The group velocity $v_g:= \partial ({\rm Re}\, \omega)/\partial k_1$ is calculated as
\begin{align}
v_g &
=\frac{1}{\sqrt{3}}{\rm sgn}(\gamma k_1)(\Gamma\gamma^2)^{1/3}t_n|k_1|^{-1/3}  \label{vg}
\end{align}
for $k_1 \neq 0$.
For a given $k_1$, the different $n$ gives the different group velocity.
The group velocity also depends on $k_1$ and it increases as $k_1$ decreases~\footnote{One might think that the group velocity (\ref{vg}) diverges at the limit $k_1 \to 0$.
However, a real system has a infrared cutoff $\sim 1/L$ where $L$ is the system size.
The ``infinite'' system size  means that $L$ is much larger than a characteristic length scale such as $\sqrt{\Gamma / \gamma}$.}.

Finally, we discuss the degeneracy of number of the gapless modes.
The dispersion approaches to zero in the limit $k_1$, $k_3 \to 0$ for any $n$.
This means the infinite number of NG modes exists.
One might think the infinite degeneracy causes a singularity. 
This will not be  the case. The number of NG modes under the shear flow is finite in a unit volume because the eigenfunction is a localized function as shown in Fig.~\ref{fig:eigenfunction}.
We note that the same phenomenon, the infinite degeneracy of zero modes but the finite mode number in the unit volume, is observed in massless Dirac particles in the presence of a magnetic field.
As a concrete example, we consider a graphene under a magnetic field $B$, whose Hamiltonian is given as~\cite{Landaulevel}
\begin{align}
H&=v_{\rm F}
\begin{pmatrix}
- \bm{\sigma}^* \cdot \bigl(\bm{p}- \frac{e}{c} \bm{A} \bigr) & 0 \\
0 & \bm{\sigma} \cdot \bigl(\bm{p}- \frac{e}{c} \bm{A} \bigr)
\end{pmatrix},
\end{align}
where $v_{\rm F}$ is the Fermi velocity of the graphene electron, $\bm{\sigma}=(\sigma_1, \sigma_2)$ the Pauli matrix, $\bm{p}$ the momentum, $e$ the electric charge, and $c$ the speed of light. We choose the Landau gauge for the vector potential $\bm{A} =(-B x_2, 0, 0)$.
Diagonalizing the Hamiltonian, we obtain the energy eigenvalues as $\epsilon_n= \text{sgn}(n)\sqrt{2 \hbar e B|n|/c}$,
where $n$ is an integer, $n=0, \pm1, \pm2,...$. In particular, the state with $n=0$ is the gapless mode.
The Hamiltonian commutes with the momentum $p_1$ and the eigenvalue of each Landau level is independent of $p_1$, so that the state is infinitely degenerate, which is similar to our results.
It is known that the number of zero modes in a unit volume is finite,
and they do not lead to a singular behavior of a thermodynamic quantity there. 

\emph{Concluding remarks.}---
\label{sec:conclusion}
We compare our results with the NG modes in isolated and open systems without the flow~\cite{Minami:2018oxl}.
In an isolated system, the type-A NG modes associated with $O(N) \to O(N-1)$ have a linear dispersion $\omega \sim k$. In an open system, it turns to diffusion modes $\omega\sim -\ri k^2$. The number of the modes in both systems is equal to the number of broken symmetries, dim$(O(N)/ O(N-1))$.
In contrast, under the shear flow, the diffusion mode turns to the fractional dispersion and the number of the modes changes to infinite.  

In other words, the diffusion mode splits into the infinite number of the fractional modes by the shear flow. 
We call the splitting modes as the rainbow NG modes by making an analogy to the rainbow splitting sunlight by a reflection at a surface of a raindrop.
We also note that the Airy function is first developed to describe the intensity of the light in neighborhood of a caustic such as the rainbow~\cite{airy1838intensity}. 

The rainbow NG modes do not belong to the classification in~Ref.~\cite{hidaka2020spontaneous}, where homogeneity is assumed. The shear flow explicitly breaks the homogeneity, and it drastically changes the behaviors of NG modes.
It is a challenging work to establish the NG theorem for systems without homogeneity such as under a nonequilibrium steady flow.

We also compare our results to dynamics of a magnetization with a constant magnetic field gradient, where the external field simply violates the homogeneity.
The equation of motion, called the Bloch-Torrey equation~\cite{torrey1956bloch, herberthson2017dynamics}, is given as 
\begin{align}
\partial_t M(t,\bm{x})=\bigr(-\ri\gamma_G g x_2 +\Gamma \nabla^2 \bigl)M(t,\bm{x}).
\label{eq:BT eq}
\end{align}
Here, $M(t,\bm{x})$ is the transverse magnetization, $\gamma_G$ the gyromagnetic ratio,  $g$ the magnetic field gradient, and $\Gamma$ the diffusion constant.
The Bloch-Torrey equation is almost the same as  Eq.~(\ref{eigeneq}).
However, $x_2$ does not couple to the wavenumber $k_1$ but just the parameter $\gamma_G g$.
The dispersion relation of (\ref{eq:BT eq}) is calculated as 
\begin{align}
\omega =&  \frac{\sqrt{3}}{2}{\rm sgn}(g)(\Gamma \gamma_G^2)^{1/3}t_n |g|^{2/3}-\frac{\ri}{2}(\Gamma \gamma_G^2)^{1/3}t_n|g|^{2/3} 
\nonumber \\
&-\ri\,\Gamma(k_1^2+k_3^2),
\end{align}
where $-t_n$ is the zeros of the Airy function.
There are the two differences.
First, the dispersion is not gapless, $\omega \neq 0$ at $k_1, k_3 \to 0$ for any $n$.
Second, there is no infinite degeneracy at $k_1 \to 0$ because it does not have the cross terms of $t_n$ and $k_1$.
Therefore, the coupling between $x_2$ and $k_1$ due to the shear flow is important 
for the remarkable behaviors of the rainbow NG modes.

The shear flow is realized as a steady flow by  balance of an external drive and a shear viscosity.
The existence of the shear flow means that the system has them and is a nonequilibrium open system. 
However,  we cannot directly see them in our model.
We originally have the equation motion of the velocity (or the momentum) field $\bm{v}$ in addition to that of $\phi^a$.
In that equation, we can explicitly see them, and it is shown in Ref.~\cite{OnukiKawasaki} that $\phi^a$ and the fluctuation of $\bm{v}$ are decoupled around the steady shear flow state.
Then, we have omitted the equation of motion of $\bm{v}$ at first in this letter.

%We expect that the nonlinear terms of Eqs.~(\ref{sigma}) and (\ref{NG}) will not change the our results by the mean field approximation. In a critical fluid under the shear flow, it is known that the critical exponents are the same as those by the mean field even if we renormalize  nonlinear terms~\cite{OnukiKawasaki}. 

We expect that the fractional exponent $2/3$ will not change in other boundary conditions.
A realistic system has the two boundaries at $x_2=0$ and $L$.
The fractional exponent is obtained from the scaling analysis, 
and it is independent of the boundary conditions. 

The fractional behavior $k_1^{2/3}$ is recently observed by the numerical simulation of an $O(2)$ model under shear flow~\cite{nakano2020long}. 
The numerical study shows that the equal-time correlation  of the steady state in $(1+2)$ dimensions behaves as
\begin{align}
\langle \pi^b(k_1,k_2=0, t) \pi^b(-k_1,k_2=0,t) \rangle \sim k_1^{-2/3}.
\end{align}
We note that the equal-time correlation at the steady state has the translational symmetry even under the shear flow~\cite{OnukiKawasaki} thanks to the Galilei symmetry of Eq.~(\ref{eq:modelA}), and here $x_2$ is Fourier transformed.
Furthermore, as a consequence of the exponent $2/3$ which is smaller than $2$, it is numerically shown that the infrared divergence of the two dimensional system is removed and the long-range order emerges there.

There are several directions for future work.
One direction is to study NG modes under heat flow. The heat flow in a fluid changes a long-range behavior of a spatial correlation~\cite{dorfman1994generic}. We expect that the flow also changes the behaviors of NG modes.

The another direction is to find a realistic system that exhibits spontaneous symmetry breaking under the shear flow.
In the real system,  the shear flow is realized by moving two adjacent walls  of the system each other.
It is then interesting to study a magnetic fluid  confined by the two moving walls. 
The magnetic fluid is a fluid where we can make the shear flow, and the spin rotational symmetry is spontaneously broken in the ordered phase. 
We leave the study on the magnetic fluid as future work.

\emph{Acknowledgements.}---
We are grateful to  Shin Nakamura, Muneto Nitta, Yukinao Akamatsu, Yuya Tanizaki, Masaru Hongo, and for useful discussions. 
YM is supported by NSF of China (Grants No. 11975199 and No. 11674283).
HN is supported by JSPS KAKENHI (Grant No. 17H01148).
YH is supported by JSPS KAKENHI (Grant No. 17H06462 and 18H01211).

\bibliographystyle{apsrev4-1}
\bibliography{ssb}

\end{document}